# Title: Strain Tuning Three-state Potts Nematicity in a Correlated Antiferromagnet

**Authors:** Kyle Hwangbo[1†], Elliott Rosenberg[1†], John Cenker[1], Qianni Jiang[1], Haidan Wen[2], Di Xiao[3,1,4], Jiun-Haw Chu[1*], Xiaodong Xu[1,3*]

[1]Department of Physics, University of Washington, Seattle, Washington 98195, USA
[2]Advanced Photon Source, Argonne National Laboratory, Lemont, Illinois 60439, USA
[3]Department of Materials Science and Engineering, University of Washington, Seattle, Washington 98195, USA
[4]Pacific Northwest National Laboratory, Richland, Washington, USA

†These authors contributed equally to this work.
*Correspondence to: xuxd@uw.edu and jhchu@uw.edu

**Abstract:** Electronic nematicity, a state in which rotational symmetry is spontaneously broken, has become a familiar characteristic of many strongly correlated materials. One widely studied example is the discovered Ising-nematicity and its interplay with superconductivity in tetragonal iron pnictides. Since nematic directors in crystalline solids are restricted by the underlying crystal symmetry, recently identified quantum material systems with three-fold rotational ($C_3$) symmetry offer a new platform to investigate nematic order with three-state Potts character. Here, we report reversible strain control of the three-state Potts nematicity in a zigzag antiferromagnetic insulator, $FePSe_3$. Probing the nematicity via optical linear dichroism, we demonstrate either $2\pi/3$ or $\pi/2$ rotation of nematic director by uniaxial strain. The nature of the nematic phase transition can also be controlled such that it undergoes a smooth crossover transition, a Potts nematic transition, or a Ising nematic flop transition. Further elastocaloric measurements demonstrate signatures of two coupled phase transitions, indicating that the nematic phase is a vestigial order arose from the antiferromagnetism. The ability to tune the nematic order with in-situ strain further enables the extraction of nematic susceptibility, which exhibits a divergent behavior near the magnetic ordering temperature that is corroborated with both linear dichroism and elastocaloric measurements. Our work points to an active control approach to manipulate and explore nematicity in three-state Potts correlated materials.

**Main text**

Electronic nematicity has drawn significant interest over the years due to its intertwined relationship with many strongly correlated phases. The importance of nematic order was highlighted in tetragonal iron-pnictide superconductors, where speculation continues that the fluctuations of the symmetry-breaking phase could be the driving pairing mechanism[1-3]. Recently, there have been increasing reports of nematic order in quantum materials with rich phase diagrams, such as twisted graphene systems[4-7] and kagome lattice superconductors[8-11], further hinting at the important role of nematicity. However, in contrast to the widely studied Ising-nematicity ($\mathbb{Z}_2$) in tetragonal iron pnictides, the nematic order in these recent systems of interest is frequently represented by a three-state Potts (TSP) model ($\mathbb{Z}_3$), reflecting their $C_3$- and $C_6$- rotational crystal symmetries. Understanding the behavior of TSP nematicity is critical in untangling its complex interplay with other strongly correlated phases.

TSP nematic order has been theorized to have properties that are distinct from Ising-nematicity[12]. While anisotropic strain always smears out an Ising nematic transition and turning it into a crossover transition, a TSP nematic transition can be switched to either a crossover transition or an Ising nematic flop transition, depending on the sign of the strain. Previous experimental investigation of TSP nematic order only investigated the re-orientation of nematic domains with static strain well-below the transition temperature[13,14], but the effect of strain near and above the phase transition has not been explored. Here we use optical linear dichroism and elastocaloric measurements in conjunction with in-situ, tunable uniaxial strain – the conjugate field to nematicity – to control the three-state Potts nematic transition and probe its associated susceptibility in a van der Waals zigzag antiferromagnet.

The material of interest, FePSe$_3$, belongs to a family of transition metal phosphorous trichalcogenide antiferromagnets ($MPX_3$, $M$ = Fe, Mn, Ni and $X$ = S, Se). In particular, FePSe$_3$ is a zigzag antiferromagnet with a Néel temperature ($T_N$) of ~108 K[15-17]. The magnetic moments point out-of-plane and mostly localize around the $Fe$-atoms, which form a honeycomb lattice with $C_3$-symmetry (Fig. 1a). This leads to three possible arrangements of the zigzag spin chains on the honeycomb structure (Fig. 1b). Below the transition temperature, the spontaneous formation of the zigzag spin chains lowers the rotational symmetry from three-fold to two-fold. This leads to anisotropy in the optical conductivity and enables optical probing of nematicity and zigzag antiferromagnetic order[18-20]. Note that the optical anisotropy is absent in the Mn-compounds of the family[19], which exhibit Néel-type spin textures. Thus, previous studies of the antiferromagnetic order in Mn-compounds have instead relied on second-harmonic generation measurements[21-23].

We first performed a phase modulation-based optical linear dichroism (LD) measurement on a thin-bulk flake (~30 nm) that was exfoliated on a SiO$_2$/Si substrate. Figure 1c shows the measured LD as a function of the incident light polarization angle ($\theta$) at different temperatures. The photon energy is chosen to be 1.95 eV as it gives rise to a large LD signal (Extended Data Fig. 1). The measured LD follows a $\cos(2\theta)$-dependence below $T_N$, which is expected if one AFM wavevector is dominant. The polarization angle dependence allows us to determine the direction of the zigzag spin chain. The most negative(maximum)-value of LD corresponds to the incident light polarization parallel (orthogonal) to the zigzag spin chain[19]. This LD measurement is similar to previous studies on NiPS$_3$ and FePS$_3$, where the concomitant rise of a strong LD signal with the onset of the zigzag AFM order was used as a proxy measure of the zigzag antiferromagnetic order [18-20]. Nevertheless, unlike the magnetic order parameter that breaks the time reversal symmetry, the LD signal probes rotational-symmetry breaking which is even under time reversal. This implies that the LD signal is proportional to the quadratic terms of zigzag antiferromagnetic order parameter, which can be non-zero due to antiferromagnetic fluctuations even in the absence of long-range magnetic order.

Figure 1d presents the temperature dependence of the maximum LD signal, which follows a trend resembling that of an order parameter for a second order phase transition. However, unlike a magnetic order parameter which should completely disappear above $T_N$, there is a persistent LD signal significantly above the Néel temperature, forming a tail-like feature. Furthermore, polarization dependent LD measurements from 110 to 150 K (Fig. 1c) clearly demonstrate the surviving $\cos(2\theta)$-dependence above $T_N$. This peculiar behavior has been previously observed in ionic crystal systems such as NiO[24] and dubbed the *fluctuation tail* in literature[25,26]. In the case of FePSe$_3$, the zigzag antiferromagnetic order vanishes above the Neel temperature, accompanying strong spin fluctuations. Since the LD has the same symmetry as a nematic order parameter, i.e., changing sign under 90-degree rotation, the observed residual LD signal suggests that there is a finite nematic order parameter far above $T_N$. This nematic order parameter is a vestigial order formed by the fluctuations of zigzag AFM order. The tail-like nature of the signal likely arises from a symmetry-breaking strain that chooses one nematic domain, coupled with the system's diverging nematic susceptibility towards $T_N$, indicative of a nematic instability in FePSe$_3$.

Similar fluctuation tails in the LD temperature dependence have been observed previously in FePS$_3$ and NiPS$_3$[18-20], but it is important to note that the sulfide compounds have a *C2/m* monoclinic structure. Therefore, the monoclinic tilt acts as the symmetry breaking mechanism, which creates a preferential direction for the zigzag spin chain. In contrast, FePSe$_3$ belongs to the *R3̄*-space group and reportedly lacks a monoclinic distortion (Fig 1a) [27]. Consequently, the three orientations of the zigzag spin chains should be energetically degenerate and the fluctuations of the nematic order between the three states should average out above $T_N$, leading to an isotropic optical response and the absence of the fluctuation tail. As the sample is cooled below $T_N$, the zigzag order should then stochastically select one of the three possible spin chain directions, leading to an archetypal first order, Potts nematic transition[13]. However, the zigzag AFM direction (i.e., the maximum LD direction) and the observed fluctuation tail are robust against various thermal cycles, implying the presence of a certain symmetry breaking mechanism that lifts the degeneracy. Possible explanations for this include small strains introduced during the exfoliation of the flakes onto the SiO$_2$/Si wafer or as-yet-unreported small monoclinic distortions in the crystal that reduce its rotational symmetry.

To gain further insight into the nematic order of FePSe$_3$, we utilized a homebuilt cryo-strain device capable of applying in-situ strain to 2D materials[28] (see Methods). The FePSe$_3$ flake was carefully aligned so that the crystal zigzag axis was parallel to the strain direction (Extended Data Fig. 2). We use the silicon Raman mode to calibrate the applied strain (See Methods, and Extended Data Fig. 3). The polarization angle dependent LD signal was then measured as a function of applied strain at base temperature (15 K), as shown in the left panel of Fig. 2a. Remarkably, the LD signal underwent a dramatic evolution as the strain is swept from compressive to tensile. The corresponding polar plot (right panel of Fig. 2a) tracks the rotation of the nematic director, which was inferred from the polarization-dependent LD response, as strain was applied.

When comparing the LD response at the two extremes of compressive and tensile strain, we found that the nematic director is separated exactly by $2\pi/3$ and the LD strengths remained comparable.

Figure 2b presents measurements conducted on the same sample as Fig. 2a, but for a different thermal cycle. The LD signal exhibits a strikingly different behavior. As shown in the polar plot, by varying the strain from being compressive to tensile, the lobe along the horizontal axis slowly shrinks, vanishes, and then grows along the orthogonal direction. There is an effective $\pi/2$ rotation in the nematic director between the two end points of compressive and tensile strain. The LD strength is also distinctly less under tensile strain than under compressive strain.

To understand the underlying mechanism, we modelled the strain-controlled population of the $\mathbb{Z}_3$-nematic domains based on two different premises. We hypothesized that the domains are determined not only from the estimated induced strain, but also from its vertical relaxation. This hypothesis was supported by our spatial LD mapping as a function of strain (see Extended Data Fig. 4). The observation presented in Fig. 2a can be reproduced with the interplay between two nematic domains with $2\pi/3$ rotation. Specifically, the LD signals of the two individual domains are described by $\cos(2\theta)$ and $\cos(2(\theta + 2\pi/3))$, respectively. The total LD signal is then a mixture of the two domains,

$$LD(\theta) = f(\varepsilon)\cos(2\theta) + [1 - f(\varepsilon)]\cos(2(\theta + 2\pi/3)),$$

where $f(\varepsilon)$ represents the strain-dependent proportion of each nematic director which is swept from 0 to 1. This domain population model describes a situation where the strain can be used to control the weight of two nematic domains that are separated by $2\pi/3$ (Fig. 2e). Figure 2c plots the simulated overall LD signal as the weight of domain is varied by strain. The excellent agreement between the simulation and the observation in Fig. 2a support our strain-controlled nematic domain population model.

To explain the trend observed in Fig. 2b, we included the third nematic domain, $\cos(2(\theta - 2\pi/3))$ in the domain population model. The overall LD signal is expressed as

$$LD(\theta) = f(\varepsilon)\cos(2\theta) + \left(\frac{1}{2}\right)[1 - f(\varepsilon)]\cos\left(2\left(\theta + \frac{2\pi}{3}\right)\right) + \left(\frac{1}{2}\right)[1 - f(\varepsilon)]\cos\left(2\left(\theta - \frac{2\pi}{3}\right)\right).$$

As shown by the cartoons in Fig. 2f, this model describes a situation where the nematic domains described by $\cos(2(\theta + 2\pi/3))$ and $\cos(2(\theta - 2\pi/3))$ are degenerate and equally populated as tensile strain is applied. Similarly, using this domain population model, we accurately obtain the trend that was observed in Fig. 2b. Under compressive strain, the zigzag spin chain aligns parallel to the strain direction as the bond distance between *Fe-Fe* atoms becomes distinctly the shortest. This strain-induced anisotropy relieves the degeneracy between the three nematic directors, and the zigzag nematic chain aligns along the shortest *Fe-Fe* bonds.

Under tensile strain, rather than having a clear direction of *Fe-Fe* bonds that are the shortest, two chains of *Fe-Fe* bonds (rotated by $2\pi/3$ between each other) are equivalent. These two zigzag chains remain energetically degenerate, and the LD response becomes a mixture of these two domains. In this case, the nematic director seems to rotate by $\pi/2$ since the vector summation of the nematic directors result in an overall vector that points orthogonally to the nematic director of the compressive side. The seamless agreement between the observed strain responses below $T_N$ and the domain population model demonstrates that strain can be used to control the orientation of the AFM zigzag spin chains. We corroborate our results further by conducting polarization-resolved Raman spectroscopy on a sample that exhibited the $2\pi/3$-shift of the nematic director (Extended Data Figs. 5 and 6). Note that strain-induced, nematic domain re-population was also observed in $Fe_{1/3}NbS_2$, which is a three-state Potts system but with triangular crystal lattice[13].

The distinct behavior between the two thermal cycles likely resulted from subtle misalignments of the sample with respect to the piezo-stack poling axis, introduced during re-mounting of the sample onto the strainer apparatus for each thermal cycle. When the sample and the piezo-stack poling axis are nearly perfectly aligned, both $\theta = \frac{\pi}{3}$ and $\frac{2\pi}{3}$ domains are favored under tensile strain, which is the case shown in Fig. 2b. However, when there is a small misalignment between the substrate and the piezo-stack poling axis, one of the domains is favored and pinned, which is the case shown in Fig. 2a.

We then examined the effects of strain on the putative fluctuation tail above $T_N$ for the thermal cycle presented in Fig. 2a. This part of the study was conducted by performing a strain sweep from compressive to tensile as the sample was warmed from the base temperature. The phonon mode of the silicon substrate was used to convert the piezoelectric voltage to applied strain for each temperature (see Methods). Figure 3a shows the temperature-dependence of LD at -0.08% strain, where the fluctuation tail clearly remains. As described above, for compressive strain, there is a delineated direction for which the nematic order will form due to the strain-induced anisotropy. Thus, rather than having a distinct phase transition, there is a crossover behavior at $T_N$, as illustrated in Fig. 3g. This transition is similar to the smeared nematic transitions that have been observed previously in iron-pnictides.

In contrast, Fig. 3b shows the temperature-dependent LD at zero strain, where the fluctuation tail nearly vanishes, and the trend resembles that of a first-order phase transition[12,13]. When the sample is in an unstrained state, the fluctuation between the three possible nematic orientations above the transition temperature results in an isotropic optical response and the disappearance of LD. We note that the TSP transition in 2D is second order with a critical exponent $\beta$ of 1/9 ($\beta = d(\log(LD))/d(\log(-T_r))$) where $T_r$ is the reduced temperature). Given that the LD signal increases by 60% of the saturated value within 2 K below the transition temperature, we can put an upper bound on the critical exponent $\beta$ of 0.128, which does not rule out the possibility of a second order 2D Potts transition. These results confirm that the fluctuation tail observed earlier in the as-exfoliated crystal is a product of strain-induced nematicity in the system and indicates a large nematic susceptibility in $FePSe_3$.

Strain control of the phase transition character was further examined by tracking the nematic director by means of examining the incident polarization angle at which the LD response reaches its maximum as a function of strain. Figure 3d shows the maximal polarization as a function of temperature for -0.08% strain. Since the formation direction of the nematic order is fixed along the strain direction, there are no changes to the polarization angle as the sample is cooled below the transition temperature, providing additional evidence of the crossover behavior. Figure 3e shows the same plot for the zero-strain response. Below the transition temperature, where we observe a finite LD response, the nematic director is separated by 30 degrees with respect to its compressive strain counterpart. This suggests that there is an equal population of two nematic domains below the transition temperature. Since the strain was swept from compressive to tensile, we hypothesize that the layers of the FePSe$_3$ flake which are closer to the substrate favour the tensile nematic states, while the layers near the vacuum interface favour the compressive nematic state, as a result of the vertical strain relaxation discussed earlier. Figure 3h illustrates the orientations of the nematic states involved in this TSP transition.

Figure 3f shows the temperature-dependent polarization angle when sufficient tensile strain (0.07%) is applied to shift the zigzag direction at base temperature and to partially lift the degeneracy above the transition temperature. With tensile strain, the formation of the nematic order along the strain direction is precluded. Above the transition temperature, there are only fluctuations between the two possible nematic states. This fluctuation also results in a finite, anisotropic response that is orthogonal to the compressive strain response (Figs. 3c and 3f), akin to the base temperature behavior that described Fig. 2f. As the sample approaches the transition temperature, it undergoes an Ising-like phase transition, making a choice between the two possible nematic directions and forming a nematic order that is rotated 120 degrees from the nematic state driven by compressive strain. Thus, there is a 30 degree shift in the nematic director as the sample is cooled below the transition temperature. The sample starts with a nematic director orthogonal to that of the compressive state above the transition temperature and forms a nematic director that is rotationally separated 120 degrees from the compressive state below the transition, as illustrated in Fig. 3i. Notably, this behavior is unique to TSP systems[12,29], where the nature of the phase transition can be controlled with strain, allowing for a smeared crossover transition, a three-state Potts nematic transition, or a Ising phase transition, as shown by our experimental results.

The coupling between nematic order and anisotropic strain of the same symmetry enables us to investigate the nematic susceptibility using the linear dichroism (LD) response induced by anisotropic compressive strain. The rate of change of LD with respect to strain, $d(LD)/d\varepsilon$, is proportional to the nematic susceptibility. Figure 4 shows the strain and polarization-dependent LD responses, at select temperature points, with a distinct response in the compressive side. Extracting the LD value by fixing the polarization orthogonal to the nematic director, we observe that the nematic order couples linearly to the applied strain, as shown in Fig. 4d-f. By fitting $d(LD)/d\varepsilon$ for all temperature points, we obtain the nematic susceptibility curve in Fig. 4g. Analogous to observations in Ising-nematic systems[30], we observe a divergent behavior of the nematic order as we approach the transition temperature.

To investigate the thermodynamic response of FePSe$_3$ arising from anisotropic strain, we measured the elastocaloric effect. Elastocaloric measurements, described in detail in Ref. [31,32], probe strain derivatives of the entropy by inducing a perturbative AC strain in the sample and measuring the corresponding temperature changes in quasi-adiabatic conditions. For systems featuring Ising-nematic phases that couple bilinearly to strain, the amplitude of the relevant elastocaloric effect will be proportional to the DC strain multiplied by the temperature derivative of the nematic susceptibility for temperatures above the phase transition:

$$\frac{dT}{d\varepsilon} = \frac{T\lambda^2 \varepsilon_o}{C_\varepsilon} \frac{d\chi}{dT}$$

Where λ is the nemato-elastic coupling, $C_\varepsilon$ is the heat capacity of the sample, and $\varepsilon_o$ is the offset DC strain the sample is experiencing. However, this equation will no longer be strictly true for systems like FePSe$_3$ which exhibit three-state Potts nematicity rather than Ising nematicity. Nonetheless, for temperatures well above the phase transition, the cubic term of the nematic order parameter in the free energy can be safely ignored for small induced strains, and so the elastocaloric effect will indeed be a proxy for the nematic susceptibility in this temperature regime. At temperatures closer to the phase transition the cubic term will become more prominent, and the extracted DC strain derivative of the elastocaloric effect will have a more complex relationship with the nematic susceptibility, including an asymmetry between the tensile and compressive strain regimes.

When strain tensor elements, which do not bilinearly couple to the order parameter but instead tune the phase transition temperature $T_S$, are induced, the elastocaloric effect will produce a signal in proximity of the phase transition (regardless of the character of the phase transition):

$$\frac{dT}{d\varepsilon} = \frac{C^c_\varepsilon}{C^t_\varepsilon}\frac{dT_S}{d\varepsilon}$$

Which is proportional to the ratio of the critical part of the heat capacity $C^c_\varepsilon$ (the part which is shifted by strain along with $T_S$) to the total heat capacity of the sample $C^t_\varepsilon$, and the derivative of $T_S$ with respect to strain.

In Figure 5a, the quantity $dT/d\varepsilon_{xx}$ (approximated by dividing the amplitude of the temperature oscillations by the AC strain amplitude) is plotted, with the color scheme indicating the DC offset strain estimated to be induced before cooling from 123 K (tensile in red, compressive in blue). Not only are nematic susceptibilities and heat capacity anomalies present, but interestingly there is a double peak structure at temperatures close to the expected antiferromagnetic phase transition. This can be interpreted as signatures of two separate phase transitions which evolve differently with DC strain: The higher temperature one is a vestigial nematic transition marked by a peak which broadens into a crossover as larger strains are applied, and the lower peak arises from the AFM transition which remains relatively sharper for all strains[29]. Furthermore, the diverging feature changes sign as DC strain is changed, indicative that a nematic susceptibility-like term proportional to the DC strain is being probed by this measurement.

Figure 5b displays the strain derivative of the normalized elastocaloric effect, extracted in compressive, near zero, and tensile strain regimes separately. This quantity is determined as detailed in the caption. In the limit of zero-strain this quantity corresponds to the temperature derivative of the nematic susceptibility but is still informative to extract in different strain conditions. The extracted "susceptibilities" are quantitatively different, corroborating the linear dichroism measurements which establish the disparate nature of the phase transition and the corresponding fluctuations under different strain regimes. The susceptibility arising from tensile DC strains appears to diverge less sharply than the compressive side, potentially indicating the phase transition possesses Ising character here. This asymmetry of this extracted value is not expected in an Ising system, and only can be prevalent in systems in which odd powers of the nematic order parameter are inherently present, like in the three state Potts model.

Our results establish FePSe$_3$ as an archetypal material system in which nematicity in a three-state Potts model can be studied. We experimentally demonstrate via both optical and thermodynamic probes that the nematic behavior in such a model is distinct from that observed in typical tetragonal systems. Notably, the nature of the phase transition can be controlled with strain. Strain can be used to have the system undergo a smeared crossover transition, Potts nematic transition, or an Ising nematic flop transition, which has never been experimentally observed before. Finally, superconductivity has been reported to emerge in FePSe$_3$ as the antiferromagnetic state is suppressed by pressure[33]. An interesting future direction is to investigate whether there is an intimate relationship between electronic nematicity and superconductivity, as has been shown in iron pnictides.

# References

1. Norman, M. R. The challenge of unconventional superconductivity. *Science* **332**, 196-200 (2011).
2. Monthoux, P., Pines, D. & Lonzarich, G. Superconductivity without phonons. *Nature* **450**, 1177-1183 (2007).
3. Malinowski, P. *et al.* Suppression of superconductivity by anisotropic strain near a nematic quantum critical point. *Nature Physics* **16**, 1189-1193 (2020).
4. Rubio-Verdú, C. *et al.* Moiré nematic phase in twisted double bilayer graphene. *Nature Physics* **18**, 196-202 (2022). https://doi.org/10.1038/s41567-021-01438-2
5. Chichinadze, D. V., Classen, L. & Chubukov, A. V. Nematic superconductivity in twisted bilayer graphene. *Physical Review B* **101** (2020). https://doi.org/10.1103/physrevb.101.224513
6. Choi, Y. *et al.* Electronic correlations in twisted bilayer graphene near the magic angle. *Nature Physics* **15**, 1174-1180 (2019). https://doi.org/10.1038/s41567-019-0606-5
7. Jiang, Y. *et al.* Charge order and broken rotational symmetry in magic-angle twisted bilayer graphene. *Nature* **573**, 91-95 (2019). https://doi.org/10.1038/s41586-019-1460-4
8. Nie, L. *et al.* Charge-density-wave-driven electronic nematicity in a kagome superconductor. *Nature* **604**, 59-64 (2022). https://doi.org/10.1038/s41586-022-04493-8
9. Xu, Y. *et al.* Three-state nematicity and magneto-optical Kerr effect in the charge density waves in kagome superconductors. *Nature Physics* **18**, 1470-1475 (2022). https://doi.org/10.1038/s41567-022-01805-7
10. Jiang, Z. *et al.* Observation of electronic nematicity driven by three-dimensional charge density wave in kagome lattice $KV_3Sb_5$. *arXiv preprint arXiv:2208.01499* (2022).
11. Wu, P. *et al.* Unidirectional electron–phonon coupling in the nematic state of a kagome superconductor. *Nature Physics* (2023). https://doi.org/10.1038/s41567-023-02031-5
12. Fernandes, R. M. & Venderbos, J. W. Nematicity with a twist: Rotational symmetry breaking in a moiré superlattice. *Science Advances* **6**, eaba8834 (2020).
13. Little, A. *et al.* Three-state nematicity in the triangular lattice antiferromagnet $Fe_{1/3}NbS_2$. *Nature Materials* **19**, 1062-1067 (2020). https://doi.org/10.1038/s41563-020-0681-0
14. Zhang, H. *et al.* Cavity-enhanced linear dichroism in a van der Waals antiferromagnet. *Nature Photonics* **16**, 311-317 (2022).
15. Cui, J. *et al.* Chirality selective magnon-phonon hybridization and magnon-induced chiral phonons in a layered zigzag antiferromagnet. *Nature Communications* **14**, 3396 (2023).
16. Haglund, A. Thermal Conductivity of MXY3 Magnetic Layered Trichalcogenides. (2019).
17. Ni, Z. *et al.* Signatures of Z $\_3$ Vestigial Potts-nematic order in van der Waals antiferromagnets. *arXiv preprint arXiv:2308.07249* (2023).
18. Hwangbo, K. *et al.* Highly anisotropic excitons and multiple phonon bound states in a van der Waals antiferromagnetic insulator. *Nature Nanotechnology*, 1-6 (2021).
19. Zhang, Q. *et al.* Observation of Giant Optical Linear Dichroism in a Zigzag Antiferromagnet FePS3. *Nano Letters* **21**, 6938-6945 (2021). https://doi.org/10.1021/acs.nanolett.1c02188
20. Zhang, X.-X. *et al.* Spin Dynamics Slowdown near the Antiferromagnetic Critical Point in Atomically Thin $FePS_3$. *Nano Letters* **21**, 5045-5052 (2021). https://doi.org/10.1021/acs.nanolett.1c00870
21. Ni, Z. *et al.* Imaging the Néel vector switching in the monolayer antiferromagnet MnPSe3 with strain-controlled Ising order. *Nature Nanotechnology* (2021). https://doi.org/10.1038/s41565-021-00885-5
22. Ni, Z. *et al.* Direct Imaging of Antiferromagnetic Domains and Anomalous Layer-Dependent Mirror Symmetry Breaking in Atomically Thin MnPS 3. *Physical review letters* **127**, 187201 (2021).
23. Chu, H. *et al.* in *Phys. Rev. Lett.* Vol. 124 27601 (American Physical Society, 2020).
24. Schäfer, F. & Kleemann, W. High‐precision refractive index measurements revealing order parameter fluctuations in $KMnF_3$ and NiO. *Journal of applied physics* **57**, 2606-2612 (1985).

25  Gehring, G. On the observation of critical indices of primary and secondary order parameters using birefringence. *Journal of Physics C: Solid State Physics* **10**, 531 (1977).
26  Ferré, J. & Gehring, G. Linear optical birefringence of magnetic crystals. *Reports on Progress in Physics* **47**, 513 (1984).
27  Wiedenmann, A., Rossat-Mignod, J., Louisy, A., Brec, R. & Rouxel, J. Neutron diffraction study of the layered compounds MnPSe3 and FePSe3. *Solid State Communications* **40**, 1067-1072 (1981).
28  Cenker, J. *et al.* Reversible strain-induced magnetic phase transition in a van der Waals magnet. *Nature Nanotechnology* **17**, 256-261 (2022). https://doi.org/10.1038/s41565-021-01052-6
29  Fernandes, R. M., Orth, P. P. & Schmalian, J. Intertwined Vestigial Order in Quantum Materials: Nematicity and Beyond. *Annual Review of Condensed Matter Physics* **10**, 133-154 (2019). https://doi.org/10.1146/annurev-conmatphys-031218-013200
30  Chu, J.-H., Kuo, H.-H., Analytis, J. G. & Fisher, I. R. Divergent nematic susceptibility in an iron arsenide superconductor. *Science* **337**, 710-712 (2012).
31  Ikeda, M. S. *et al.* AC elastocaloric effect as a probe for thermodynamic signatures of continuous phase transitions. *Review of Scientific Instruments* **90** (2019).
32  Ikeda, M. S. *et al.* Elastocaloric signature of nematic fluctuations. *Proceedings of the National Academy of Sciences* **118**, e2105911118 (2021).
33  Wang, Y. *et al.* Emergent superconductivity in an iron-based honeycomb lattice initiated by pressure-driven spin-crossover. *Nature Communications* **9** (2018). https://doi.org/10.1038/s41467-018-04326-1
34  Guo, Y. *et al.* Distinctive in-Plane Cleavage Behaviors of Two-Dimensional Layered Materials. *ACS Nano* **10**, 8980-8988 (2016). https://doi.org/10.1021/acsnano.6b05063
35  Bhutani, A., Zuo, J. L., McAuliffe, R. D., dela Cruz, C. R. & Shoemaker, D. P. Strong anisotropy in the mixed antiferromagnetic system Mn 1− x Fe x PSe 3. *Physical Review Materials* **4**, 034411 (2020).

**Acknowledgements:** We thank Q. Zhang for substantial insights. This work was mainly supported by the Department of Energy, Basic Energy Sciences, Materials Sciences and Engineering Division (DE-SC0012509). Strain devices were partially supported by Air Force Office of Scientific Research (AFOSR) Multidisciplinary University Research Initiative (MURI) program, grant no. FA9550-19-1-0390. Bulk crystal growth and the elastocaloric measurements were supported by NSF MRSEC DMR-1719797, DMR-2308979 and the Gordon and Betty Moore Foundation's EPiQS Initiative, Grant GBMF6759 to JHC. The authors also acknowledge the use of the facilities and instrumentation supported by NSF MRSEC DMR-1719797. XX and JHC acknowledges the support from the State of Washington funded Clean Energy Institute.

**Author contributions:** XX, KH, JHC, and ER conceived the experiment. KH fabricated samples and performed optical measurements. JC designed and built the strain cell for atomically thin flakes. QJ and JHC synthesized and characterized the bulk crystals. DX constructed the domain population model. All authors contributed to the data analysis and interpretation. KH, XX, ER and JHC wrote the paper with input from all authors. All authors discussed the results and commented on the manuscript.

**Competing interests:** The authors declare no competing financial interests.

**Data availability:** The datasets generated during and/or analyzed during this study are available from the corresponding author upon reasonable request.

**Methods:**

**Crystal growth.** Single crystals of FePSe$_3$ were synthesized by the chemical vapor transport method using iodine as the transport agent. Stoichiometric amounts of iron powder (99.998%), phosphorus powder (98.9%), and selenium powder (99.999%) were mixed with iodine (1~mg/cm$^3$) and sealed in quartz tubes (~10cm in length) under high vacuum. The tubes were placed in a horizontal two-zone furnace. Large and thin crystals (~10×10×0.01 mm$^3$) of FePSe$_3$ were then obtained after quickly heating the precursor up to 800°C for the source end and 750°C for the sink end, dwelling for 12 hours and quickly cooling down to room temperature.

**Strain measurements and strain calibration.** To deterministically apply strain along certain directions of the *Fe*-honeycomb lattice, we first exfoliated the flakes onto PDMS slabs. We found FePSe$_3$ flakes to have distinctive cleavage direction when exfoliated onto substrates, as is the case with many other vdW materials[34], which allowed us to infer the zigzag/armchair direction of the material from the sample images. The flakes were then transferred onto thin silicon rectangles that were adhered to 2D flexure sample plates produced by Razorbill instruments using Stycast 2850 FT epoxy with the identified crystal axes carefully aligned to the strain direction. Extended Data Fig. 2 shows an optical image of the prepared strain sample. Strain measurements were conducted using home-built strain cells that are described in Ref. [28]. The strength of the applied strain to the sample was measured by tracking the Raman shift of a silicon phonon peak (center at ~525 cm$^{-1}$) as a function of the piezo voltage (Extended Data Fig. 3). The Raman shift was then converted to a strain value by using the previously reported strain shift rate. Previous study (Ref. [28]) indicates that the strain shift rate of the silicon phonon mode is an accurate measure of the applied strain to the sample. The presented strain values have been shifted to account for the built-in strain from sample preparation and thermal expansion coefficient mismatch in the strain system. The built-in strain was identified through domain population modeling of the linear dichroism response of the sample at base temperature. This study was conducted using a cold-finger cryostat from Montana Instruments. Due to the large thermal load introduced by the cryo-strainer, there is a thermal shift for the strain samples with respect to the non-strain sample presented. We have corrected for this thermal offset in the measurements presented for the strained sample. The offset was identified by measuring a non-strained FePSe$_3$ sample in the same measurement scheme and noting the shift in the transition temperature (Extended Data Fig. 3).

**Linear dichroism (LD) spectroscopy.** The measurements were carried out in the reflection geometry. A 633 nm He-Ne laser was doubly modulated by a photoelastic modulator (PEM) with a retardance of λ/2 and a mechanical chopper. After phase modulation, the light passed through a half-waveplate, and then was focused down onto the sample at normal incidence with an objective lens. A laser power of ~5 μW was used. The reflected light was detected by a photodiode and demodulated at 100 kHz and 1 kHz, which corresponds to the PEM linear polarization modulation and the chopper modulation frequency, respectively. We also conducted LD spectrum measurements using a supercontinuum laser and a tuneable filter set with 1 nm spectral resolution. We observed a broad LD response centred about 620 nm (Extended Data Fig. 1). The broad response allows 633 nm laser to be a suitable excitation energy for the measurements. Linear dichroism reading from the sample at 295 K was subtracted from the presented data to remove background anisotropy introduced by optical components. The linear dichroism measurement probes the nematic order parameter ($\boldsymbol{n}$), which is expressed in terms of the magnetic order parameters ($L_i$, where $i = 1,2,3$ and corresponds to each choice of zigzag AFM order)[13]:

$$\boldsymbol{n} = (n_1, n_2) \propto \left(|L_1|^2 + |L_2|^2 - 2|L_3|^2, \sqrt{3}|L_1|^2 - \sqrt{3}|L_2|^2\right).$$

**Raman measurement**. A HeNe laser (633 nm) was used to excite the FePSe$_3$ sample, which was placed in a closed cycle cryostat with temperature range from 5 K to 300 K. The Raman measurements were polarization-resolved and collected by a spectrometer with a liquid nitrogen cooled CCD camera (Extended Data Fig. 5 and 6).

**Elastocaloric measurements:** For the elastocaloric measurements, a commercial Razorbill CS-100 strain cell was used to apply strain to the samples, cut 90 degrees from the crystal facets to be approximately 1 mm x 0.4 mm x 0.02 mm in size. The samples were secured between two sets of mounting plates using Stycast 2850FT Epoxy, which were screwed into the strain cell, to have a gap of approximately 0.7mm. An AC voltage of 5V RMS at 14 Hz was applied to the outer piezoelectric (PZT) stacks of the strain cell, which corresponded to applying an AC displacement of the sample of approximately 0.01% of its length. This frequency was experimentally determined by measuring the elastocaloric signal at 120 K for frequencies in the range of 5-50 Hz and choosing the frequency with the largest response. This implied the frequency was at the plateau of the relevant thermal transfer function, which did not observably shift in the temperature range measured. DC voltages were applied to the inner PZT to reach a strain range of 0.7%. To approximate the strain the sample experienced a capacitor built in the strain cell was measured to provide the relative displacement of the sample plates, which was divided by the length of the gap. This however only approximated $\varepsilon_{xx}$ of the sample as it assumes a 100% strain transmission.

The temperature fluctuations in the sample induced by the AC strain were measured using a homemade Type E (Chromel-Constantan) thermocouple. The chromel and constantan wires (50 µm diameter) were thermally anchored to an outer part of the strain cell and silver pasted together to the sample, as shown in the inset of Figure 5. The voltage between the two wires was measured with an SRS860 at the frequency of the strain being applied to obtain the amplitude of the temperature fluctuations.

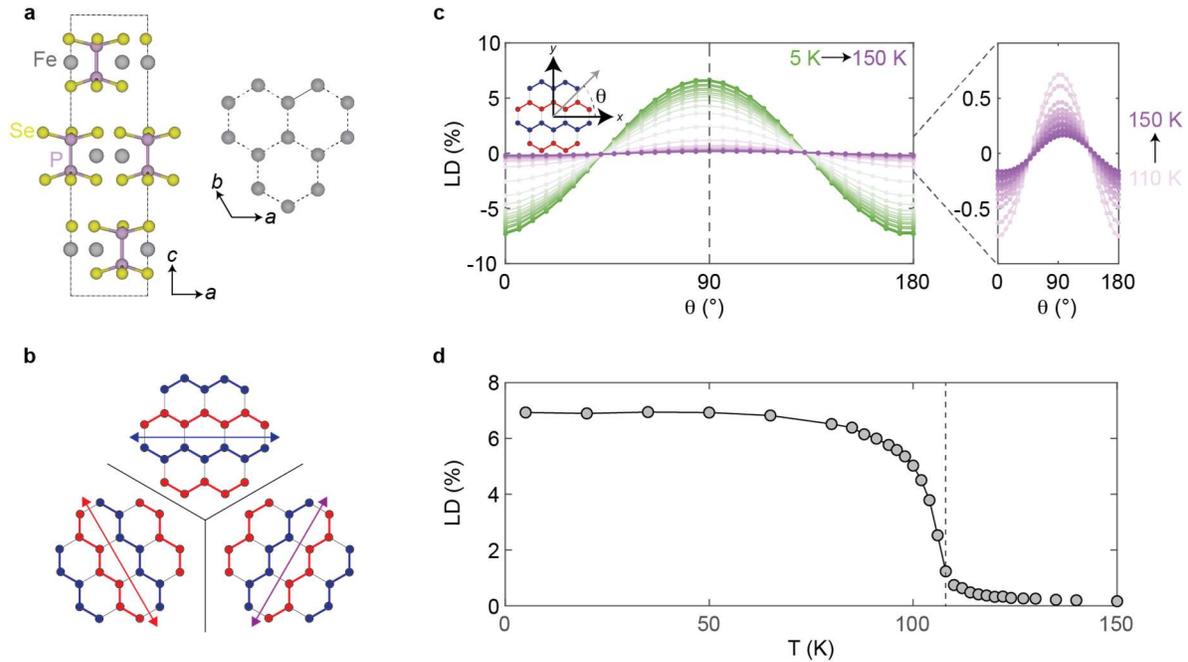

**Fig. 1 | Zigzag antiferromagnetic order and three-state degeneracy. a,** Crystal structure of FePSe$_3$. The *Fe*-atoms form a honeycomb lattice when viewed in the *ab*-plane. **b,** Three possible choices of the zigzag spin-chain directions on a honeycomb lattice. The arrows represent the nematic director corresponding to each choice of the zigzag AFM order. **c,** Optical linear dichroism (LD) response of a thin-bulk FePSe$_3$ flake as the polarization angle of the incident light ($\theta$) is rotated with respect to the zigzag spin chain direction, at various temperatures. Right panel is the Zoom-in on the polarization-dependent LD signal at temperatures above $T_N$. **d,** LD response as a function of temperature when the incident polarization is orthogonal to the zigzag spin chains (polarization angle indicated by the grey dashed line in **(c)**). The dashed line indicates the $T_N$.

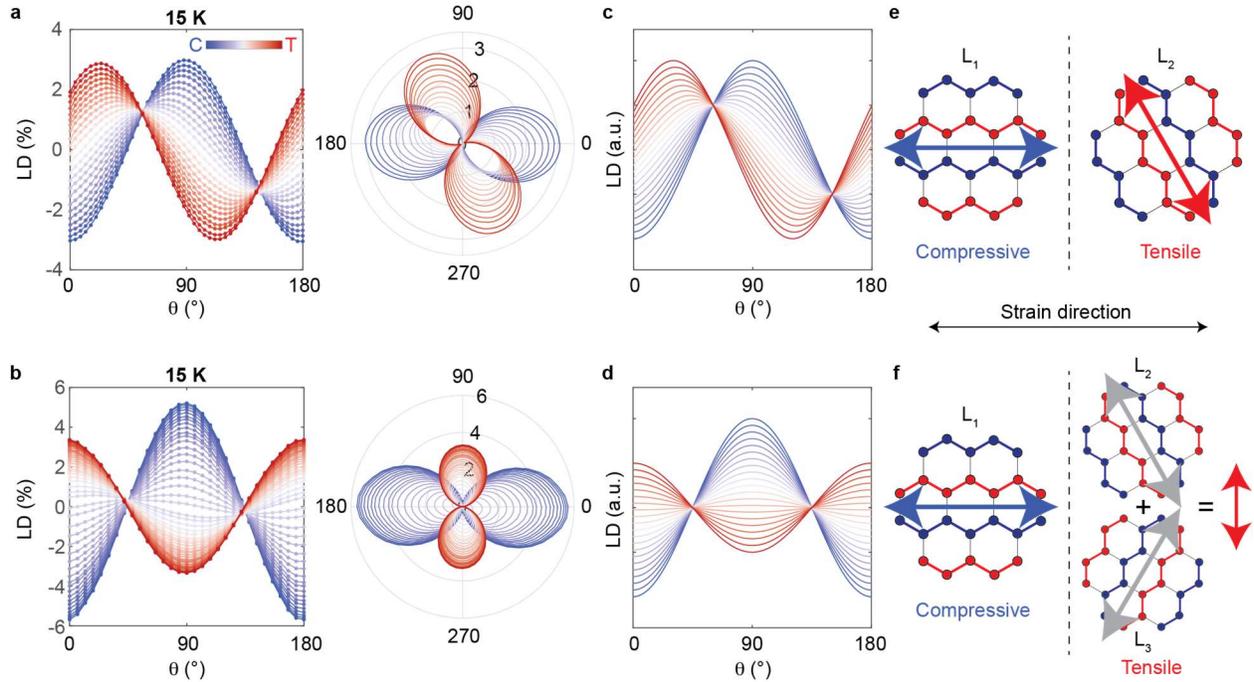

**Fig. 2 | Strain control of zigzag AFM order at 15 K. a, b,** Polarization-dependent LD response as strain is applied for two different thermal cycles. The polar graphs track the rotation of the nematic director as strain is applied. **c, d,** Simulated LD responses using nematic domain population model (see text for details). **e,** Illustration of the zigzag AFM order at maximal compressive and tensile strain applied for the sample response presented in **(a)**. The arrows represent the nematic director at the respective strain states. **f,** Illustration of the zigzag AFM order presented at maximal strain points for sample response presented in **(b)**. For the tensile response, there are coexisting zigzag AFM domains where the vector summation of the domains' nematic directors (grey arrows) result in an overall nematic director (red arrow) that is vertical (orthogonal to the nematic director of the compressive strain).

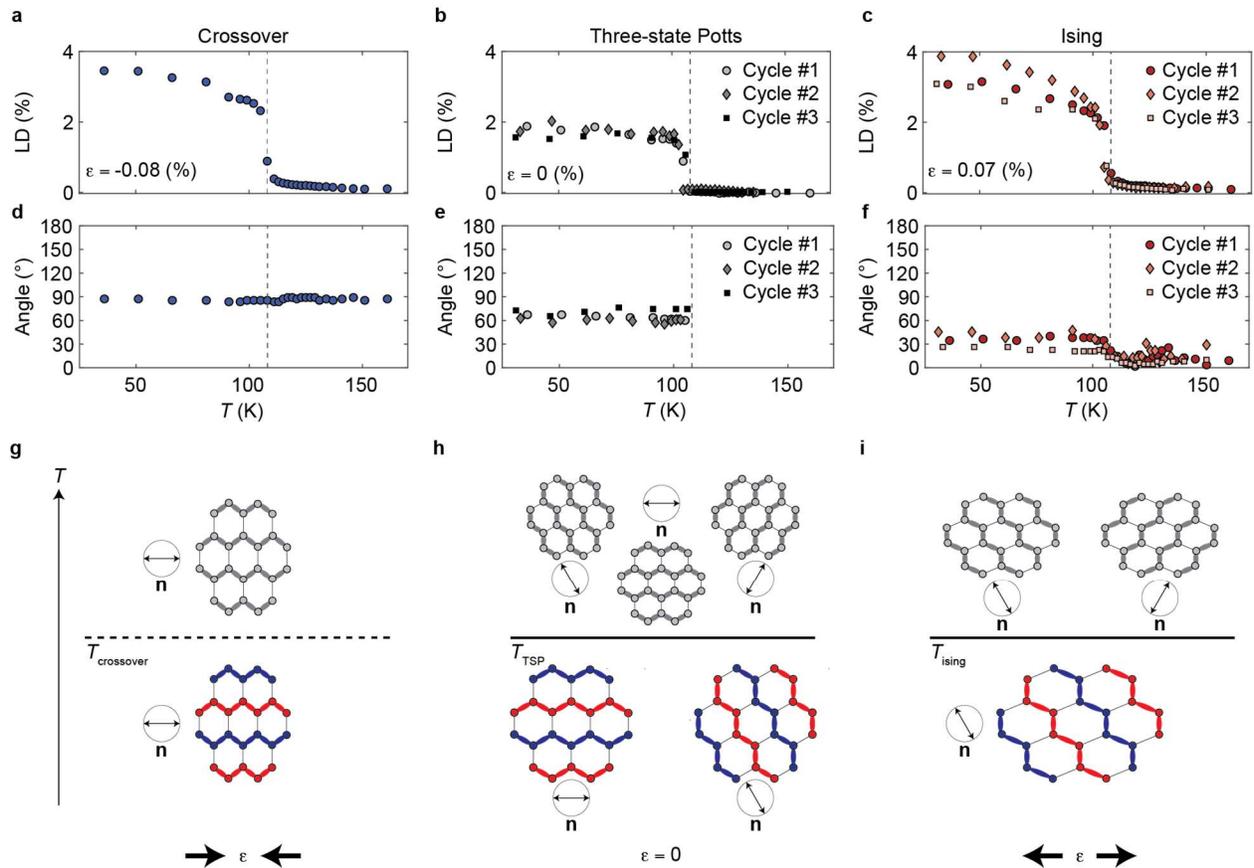

**Fig. 3 | Strain control of nematic phase transition. a-c,** LD response as a function of temperature for **(a)** compressive strain, **(b)** zero-strain, and **(c)** tensile strain. The incident polarization angle was orthogonal to the nematic director. The dashed line marks the magnetic transition temperature (~108 K). **(b)** and **(c)** show the responses for three different thermal cycles of the same sample. **d-f,** Evolution of the nematic director as a function of temperature for different strains. The angles are offset by 90 degrees at each temperature point. Points above the transition temperature is omitted for **(e)** due to the lack of LD response. **g-i,** Illustration of the nematic order as temperature is lowered with **(g)** compressive strain, **(h)** zero-strain, and **(i)** tensile strain. The thicker, grey bonds highlight the direction of the nematic order. **(g)** depicts the crossover transition that is observed in **a** and **d**, while **(h)** illustrates the three-state Potts transition presented in **b** and **e**. **(i)** depicts the Ising transition shown in **c** and **f**.

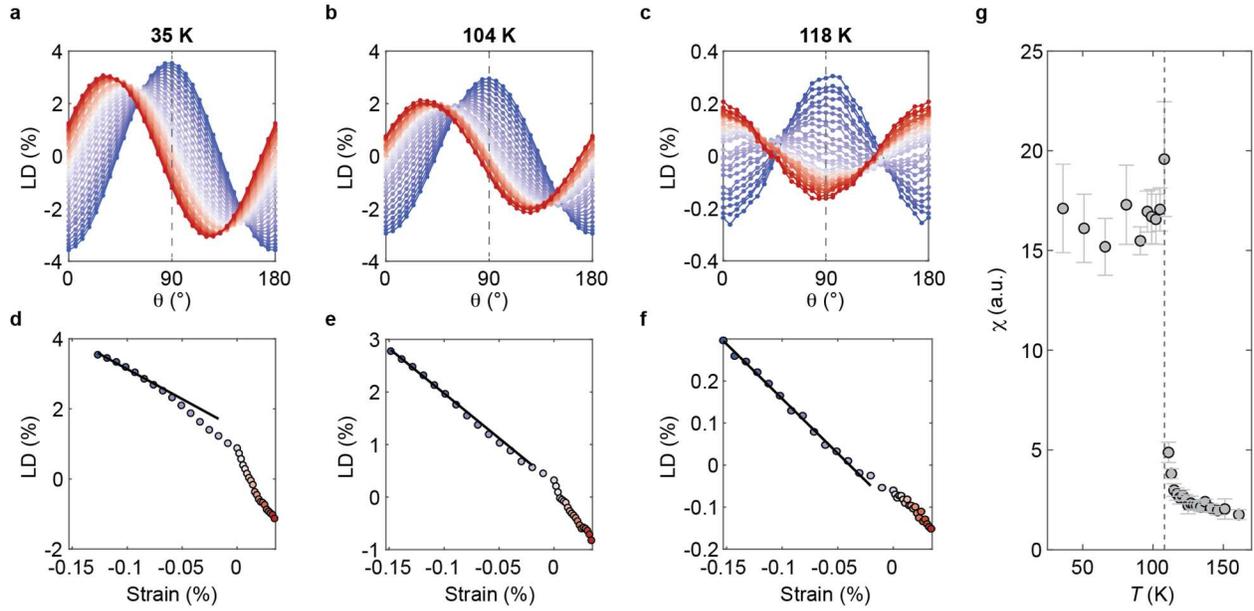

**Fig. 4 | Nematic susceptibility of the compressive side. a-c,** Polarization-dependent LD response as a function of strain at different temperatures. **d-f**, LD values at a fixed polarization angle, indicated by the dashed lines in panel **(a-c)** as strain is applied for 35 K, 104 K, and 118 K, respectively. The black line exemplifies the linear fitting that was performed for the compressive side to extract the rate of change in LD strength versus strain at different temperature points. **g,** Temperature dependence of the nematic susceptibility obtained by $d(LD)/d\varepsilon$. The dashed line indicates the transition temperature (~108 K). Error bars represent the confidence bounds of the slope values extracted from the linear line fits.

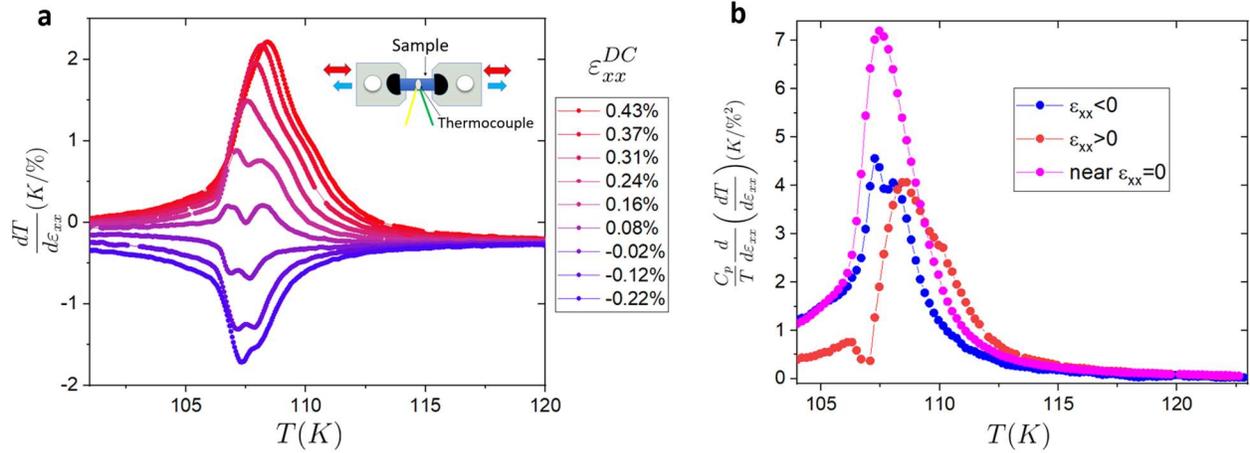

**Fig. 5 | Elastocaloric measurements of FePSe$_3$. a**, displays the measured amplitude of the temperature fluctuations divided by the amplitude of the estimated AC strain. DC strains were induced at 123 K (denoted by the color scheme, tensile is red, compressive blue), and then system was cooled to 100 K. Zero-strain was estimated by determining the DC strain where the minimum of the transition temperatures occurred and where the diverging EC signal changed concavity. **b**, displays the strain derivative of the normalized elastocaloric effect with the colors dictating what DC strain regime the linear fit was performed in, tensile (red), near zero-strain (pink), and compressive (blue). These quantities were determined by first normalizing the data in panel (a) by the heat capacity background (from Ref [35]) divided by temperature ratio, and then binning the data for select temperatures. A linear fit was performed of the binned data versus DC strain for each strain regime at each temperature, and the resulting slopes are plotted.

# Extended Data Figures for
# Strain Tuning Three-state Potts Nematicity in a Correlated Antiferromagnet

**Authors:** Kyle Hwangbo[1†], Elliott Rosenberg[1†], John Cenker[1], Qianni Jiang[1], Haidan Wen[2], Di Xiao[3,1,4], Jiun-Haw Chu[1*], Xiaodong Xu[1,3*]

[1]Department of Physics, University of Washington, Seattle, Washington 98195, USA
[2]Advanced Photon Source, Argonne National Laboratory, Lemont, Illinois 60439, USA
[3]Department of Materials Science and Engineering, University of Washington, Seattle, Washington 98195, USA
[4]Pacific Northwest National Laboratory, Richland, Washington, USA

†These authors contributed equally to this work.
*Correspondence to: xuxd@uw.edu and jhchu@uw.edu

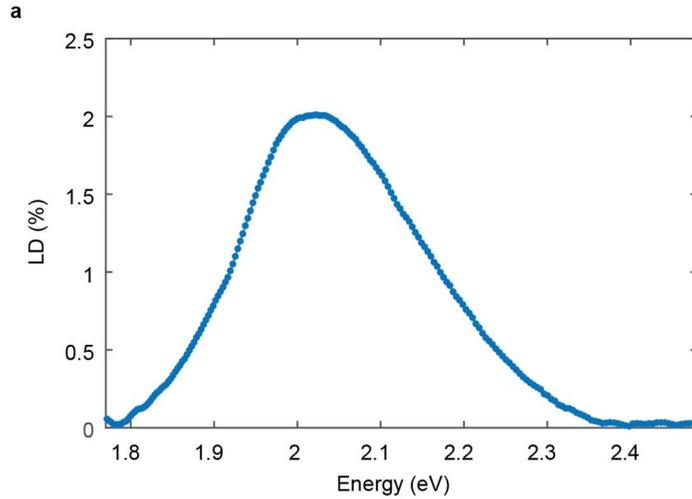

**Extended Data Fig. 1 | Linear dichroism spectrum. a,** Linear dichroism signal as a function of incident photon energy for a thin-bulk FePSe$_3$ sample at 5 K. The spectrum shows a broad LD response centered around 2 eV. The photon energy dependent LD response is possibly due to a resonance effect with a *d-d* electronic transition. Previous measurements on FePS$_3$ showed similar enhancement of LD response at known *d-d* transitions.

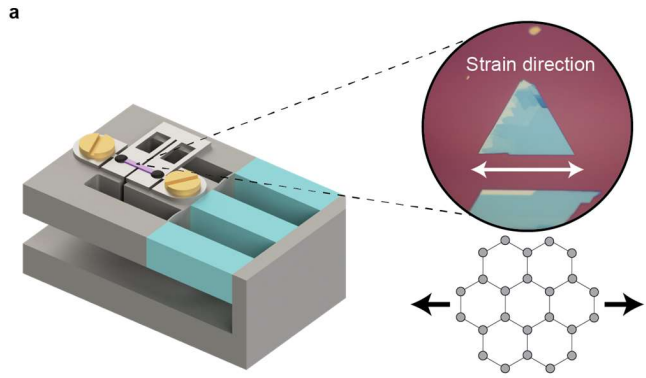

**Extended Data Fig. 2 | Strain apparatus and crystal response to strain. a,** Schematic of the strain cell used for the strain measurements and optical image of a representative strain sample. The crystal zigzag direction was aligned along the strain direction.

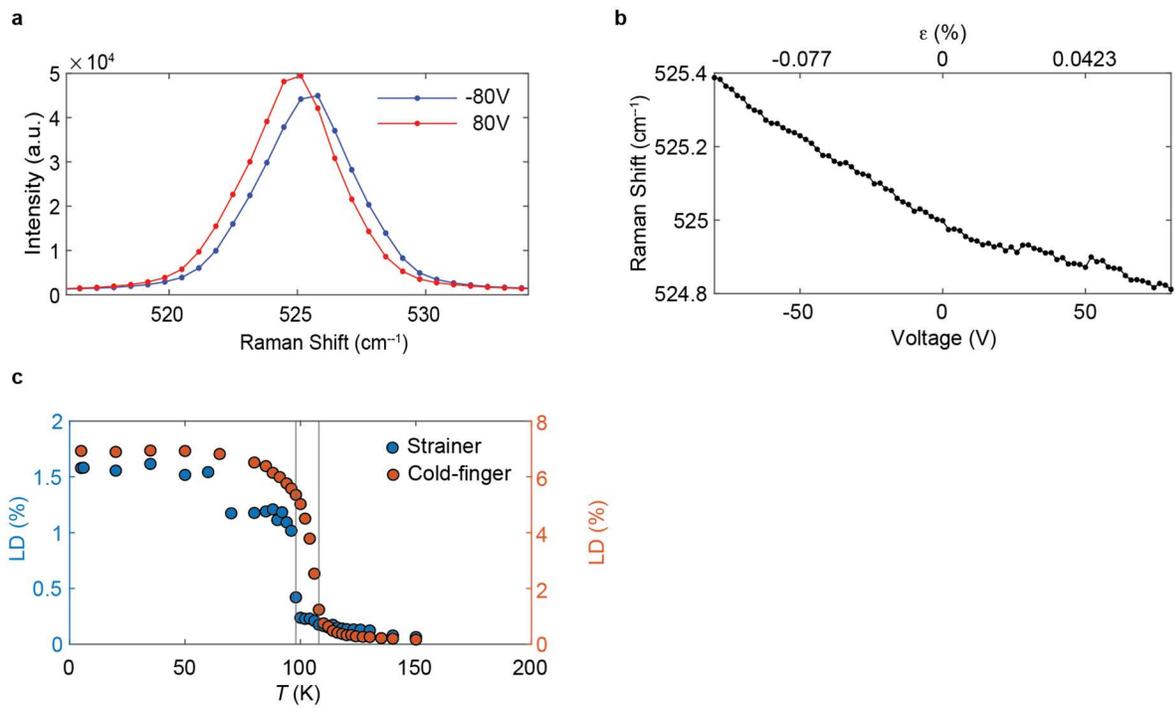

**Extended Data Fig. 3 | Strain calibration using Si phonon mode and Thermal calibration. a,** Raman spectra of the silicon Raman mode centered around 525 cm$^{-1}$ with -80 V (blue) and 80 V (red) applied to the strain cell. **b,** Raman shift of the silicon Raman mode as a function of the piezo voltage. **c,** Linear dichroism versus temperature for a non-strained, thin bulk FePSe$_3$ flake on the strainer setup, showing the shift (~ 10 K) in the transition temperature due to the increased thermal load of the strainer. The x-axis shows the nominal thermocouple reading.

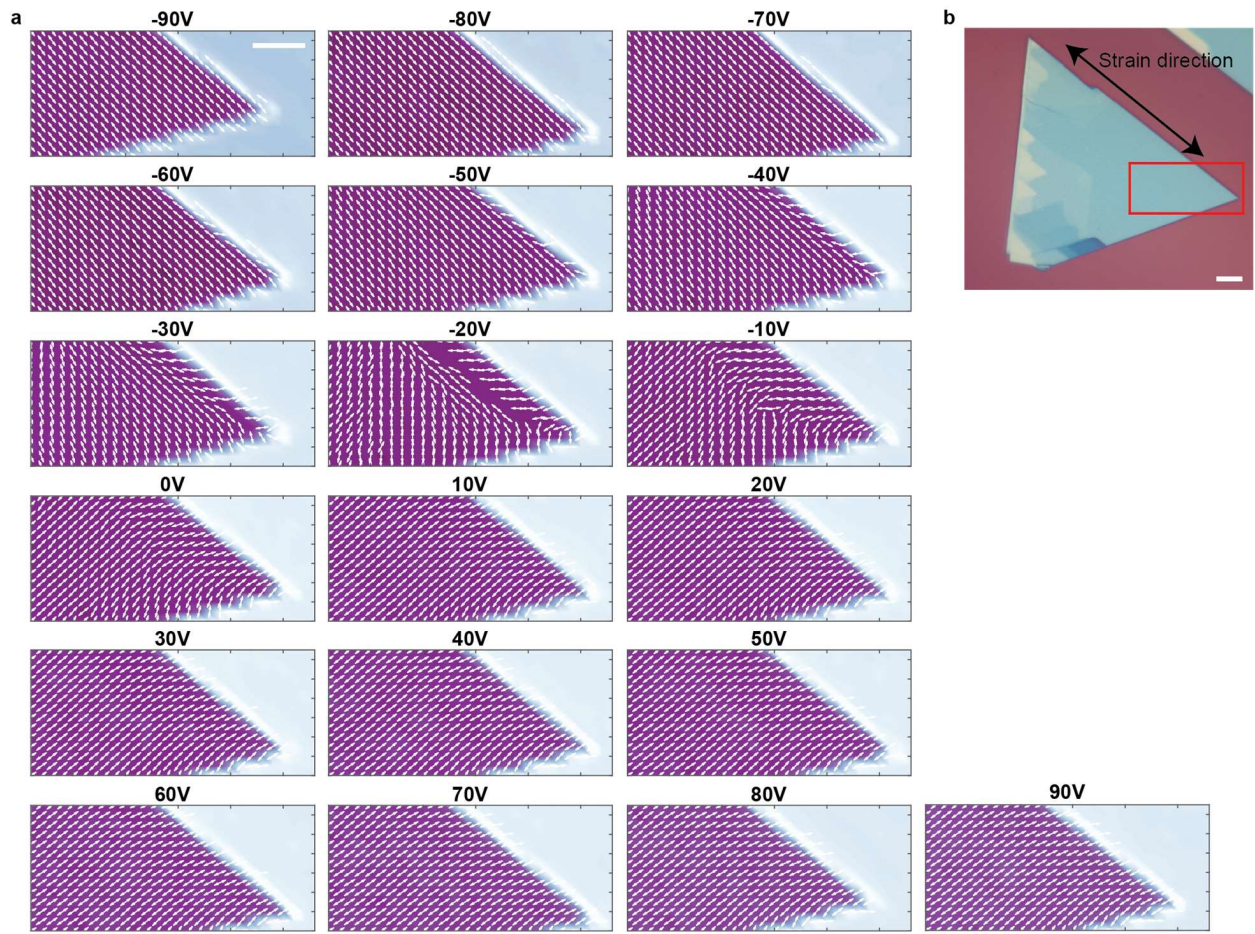

**Extended Data Fig. 4 | Linear dichroism spatial mapping. a,** Nematic directors (represented by the white arrows) at each sampled point overlaid on top of the reflection raster map for various piezo voltages. The raster map was scanned over 28 $\mu m$ x 14 $\mu m$ area with 1 $\mu m$ step sizes. The maximally negative voltage corresponds to the highest compressive strain applied. Scale bar: 5 $\mu m$. **b,** Optical image of the measured strain sample. The red box represents the area which the LD mapping was measured, and the black arrow shows the strain direction. Scale bar: 5 $\mu m$.

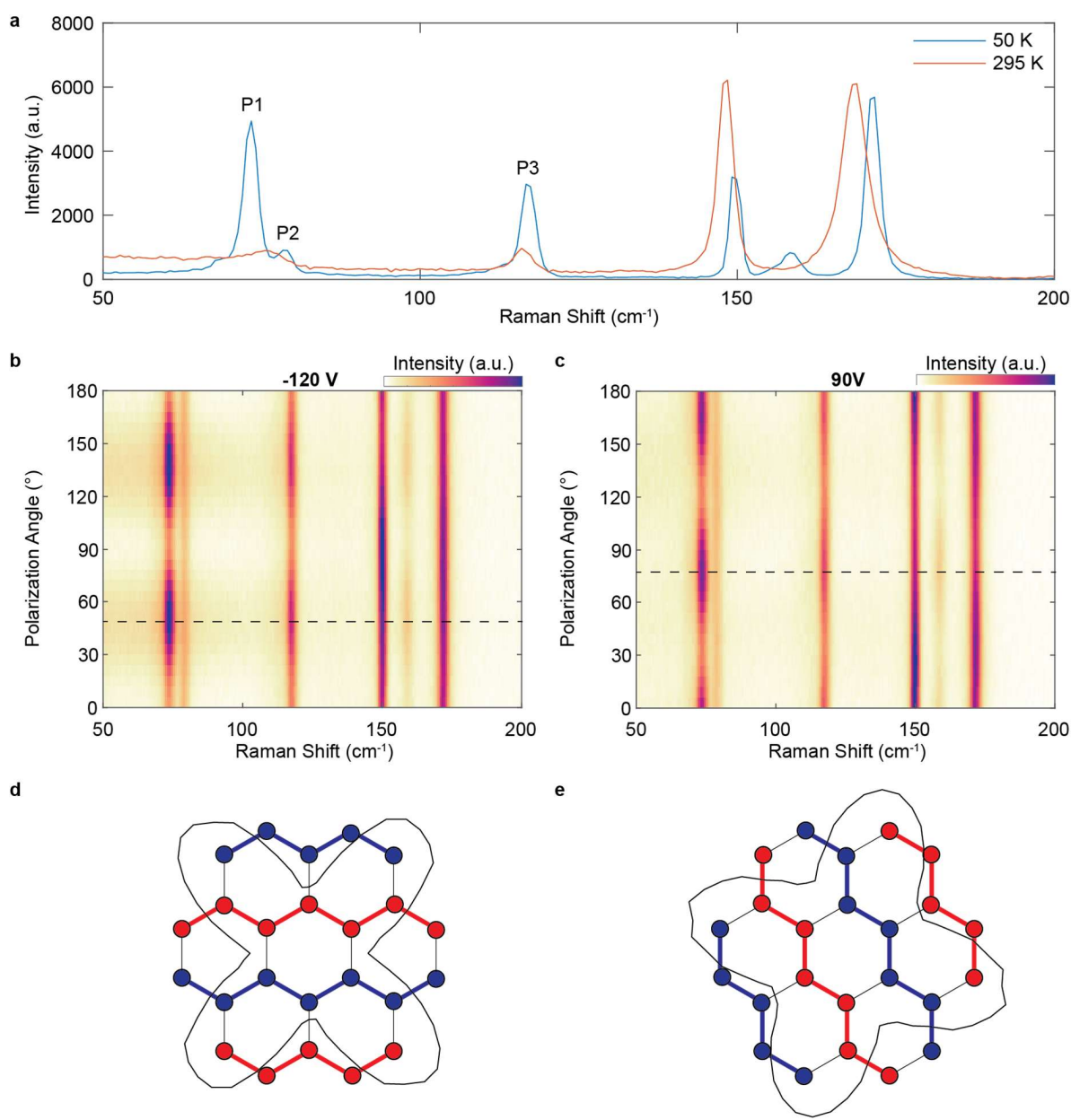

**Extended Data Fig. 5 | Polarization-resolved Raman spectroscopy of the strain FePSe$_3$ sample a,** Raman response of a thin-bulk flake at two thermal points. Labelled peaks (P1, P2, and P3) correspond to Raman modes associated with the zigzag AFM order. **b, c,** Co-linearly polarized Raman scattering as the incident polarization is rotated for compressive (-120 V) and tensile (90 V) strain, respectively. **d, e,** Polar plot of the integrated intensity of the Raman mode labelled P1 superimposed onto zigzag orders at compressive **(d)** and tensile **(e)** strain. P1 Raman mode exhibits a four-fold symmetry that is rotationally separated by $\pi/4$ with respect to the zigzag order. Thus, a $2\pi/3$-shift in the zigzag order would result in a $\pi/6$-shift in the incident polarization dependence of the Raman mode.

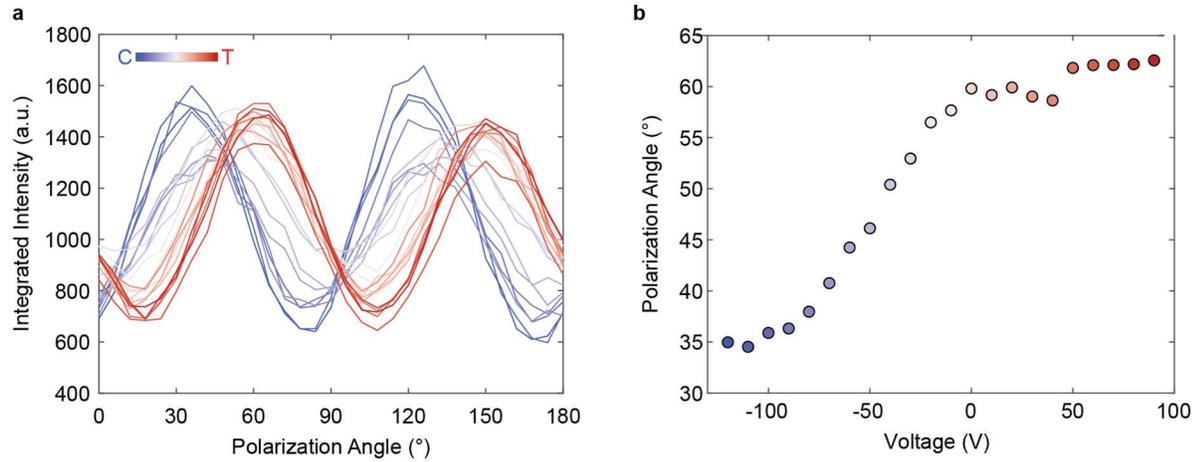

**Extended Data Fig. 6 | Polarization-dependent intensity of Raman mode at ~73 cm$^{-1}$. a,** Polarization dependence of the integrated intensity of Raman mode labelled P1 in Extended Data Fig. 5 as strain is swept from compressive (blue curves) to tensile (red curves) side. **b,** Incident polarization angle values where the Raman peak reaches the maximum value. There is near 30-degree rotation as strain is swept from compressive to tensile strain.